\documentclass[9pt, a4paper]{article}

\usepackage{a4wide}
\usepackage{xcolor}
\usepackage{amssymb}
\usepackage{amsmath}
\usepackage{graphicx}
\usepackage{url}
\usepackage{bm}
\usepackage[utf8]{inputenc}
\usepackage[parfill]{parskip}
\usepackage{xr} 
\usepackage{subcaption}
\usepackage{cleveref}
\usepackage{multirow}
\usepackage[title]{appendix}

\usepackage{soul} 

\usepackage{titling} 
\usepackage{authblk} 

\title{COFFEE: COVID-19 Forecasts using Fast Evaluations and Estimation}

\author[1,2]{Lauren Castro}
\author[1]{Geoffrey Fairchild}
\author[3]{Isaac Michaud}
\author[3]{Dave Osthus}

\affil[1]{Information Systems and Modeling Group, Los Alamos National Laboratory}
\affil[2]{Center for Nonlinear Studies, Los Alamos National Laboratory}
\affil[3]{Statistical Sciences Group, Los Alamos National Laboratory}

\date{}

\begin{document}

\maketitle

\abstract{This document details the methodology of the Los Alamos National Laboratory COVID-19 forecasting model, COFFEE (COVID-19 Forecasts using Fast Evaluations and Estimation).\footnote{Approved for unlimited release and assigned number LA-UR-20-28630}}

\section*{COFFEE Methodology}
COFFEE is a probabilistic model that forecasts daily reported cases and deaths of COVID-19. COFFEE is fit to geographic regions independently, facilitating parallelization for fast computations. 

\subsection*{Notation}

\begin{itemize}
\item $t$ indexes time, where $t$ is the number of days from a reference starting date (index)
\item $T$ is the day of the last observation (index)
\item $K$ is the forecast window size, in days (index)
\item $y_{c,t}$ is the number of reported cases of COVID-19 on day $t$ as reported on the COVID-19 Dashboard by the Centers for Systems Science Engineering (CSSE) at Johns Hopkins University (JHU) (observable)
\item $\ddot{y}_{c,t} = \sum_{j=1}^t y_{c,j}$ is the cumulative number of reported cases of COVID-19 through day $t$ as reported by CSSE at JHU (observable)
\item $y_{d,t}$ is the number of reported deaths of COVID-19 on day $t$ as reported by CSSE at JHU (observable)
\item $\ddot{y}_{d,t} = \sum_{j=1}^t y_{d,j}$ is the cumulative number of reported deaths of COVID-19 through day $t$ as reported by CSSE at JHU (observable)
\item $\delta_{c,t}$ is the underlying number of reported cases on day $t$ (unobservable)
\item $\ddot{\delta}_{c,t} = \sum_{j=1}^t \delta_{c,j}$ is the underlying number of cumulative reported cases through day $t$ (unobservable)
\item $\delta_{d,t}$ is the underlying number of reported deaths on day $t$ (unobservable)
\item $\ddot{\delta}_{d,t} = \sum_{j=1}^t \delta_{d,j}$ is the underlying number of cumulative reported deaths through day $t$ (unobservable)
\item $\delta_{s,0}$ is the underlying number of susceptible individuals at the start of the pandemic (unobservable)
\item $\delta_{s,t} = \delta_{s,0} - \ddot{\delta}_{c,t}$ is the underlying number of susceptible individuals on day $t$ (unobservable)
\end{itemize}

We use the convention that bolded quantities are vectors and unbolded quantities are scalars. For concreteness, $y_{c,t}$ is a scalar while $\boldsymbol{y}_{c,1:t} = (y_{c,1},y_{c,2},\ldots, y_{c,t})'$ is a $t \times 1$ vector.

\subsection*{Cases Model}

Let
\begin{align}
y_{c,t}|\delta_{c,t}, \alpha &\sim \text{NB}\Bigg(\delta_{c,t}, \frac{\delta_{c,t}}{\alpha}\Bigg)
\end{align}
\noindent where NB(a,b) is a Negative-Binomial model with mean parameter a $> 0$ and size parameter b $> 0$ where
\begin{align}
\text{E}(y_{c,t}|\delta_{c,t},\alpha) &= \delta_{c,t} \\
\text{Var}(y_{c,t}|\delta_{c,t},\alpha) &= \delta_{c,t}(1 + \alpha).
\end{align}

\begin{figure}[h!]
    \centering
	 \includegraphics[width=1\linewidth]{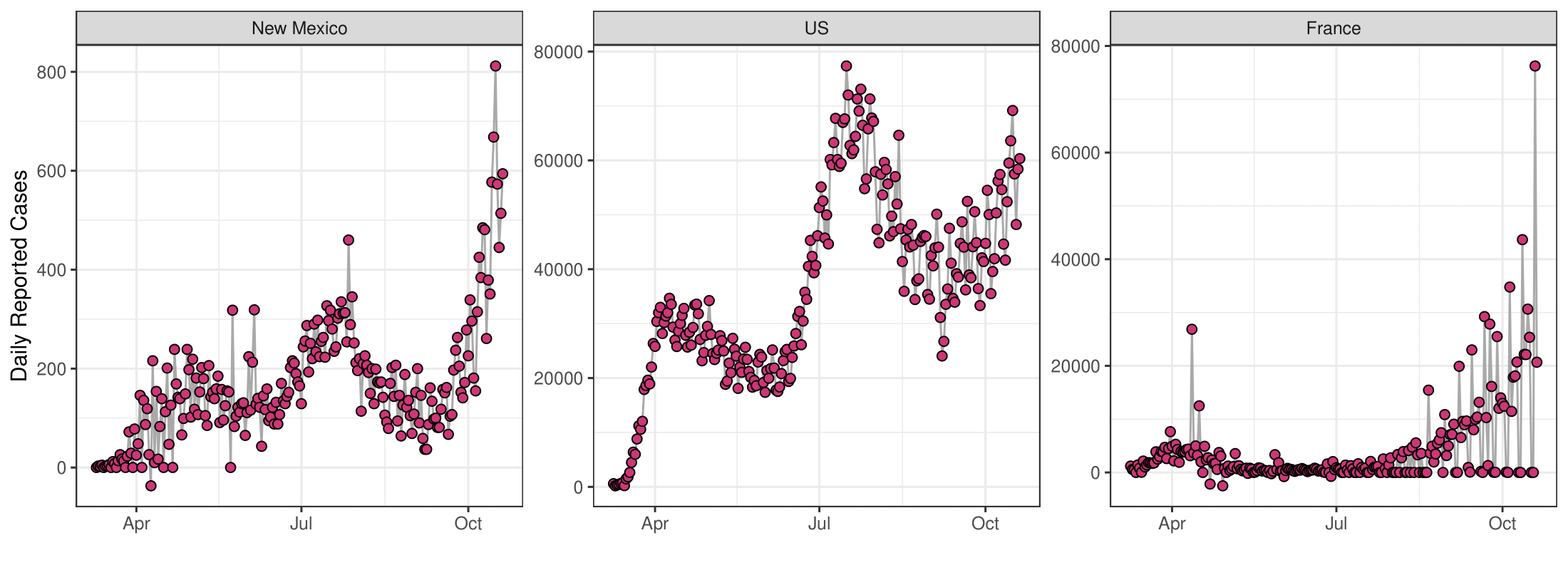}
          \caption{The daily reported cases of COVID-19 for New Mexico, the US, and France.}
          \label{fig:cases}
\end{figure}

Figure \ref{fig:cases} shows the daily reported cases for New Mexico, the United States (US), and France. All three regions have gone through rising and declining periods of cases with various levels of noise in the reported cases. 

In what follows, we outline the steps COFFEE takes to produce forecasts of reported cases.

\subsubsection*{Step 1: Identify and Adjust Outliers}
\begin{figure}[h!]
    \centering
	 \includegraphics[width=1\linewidth]{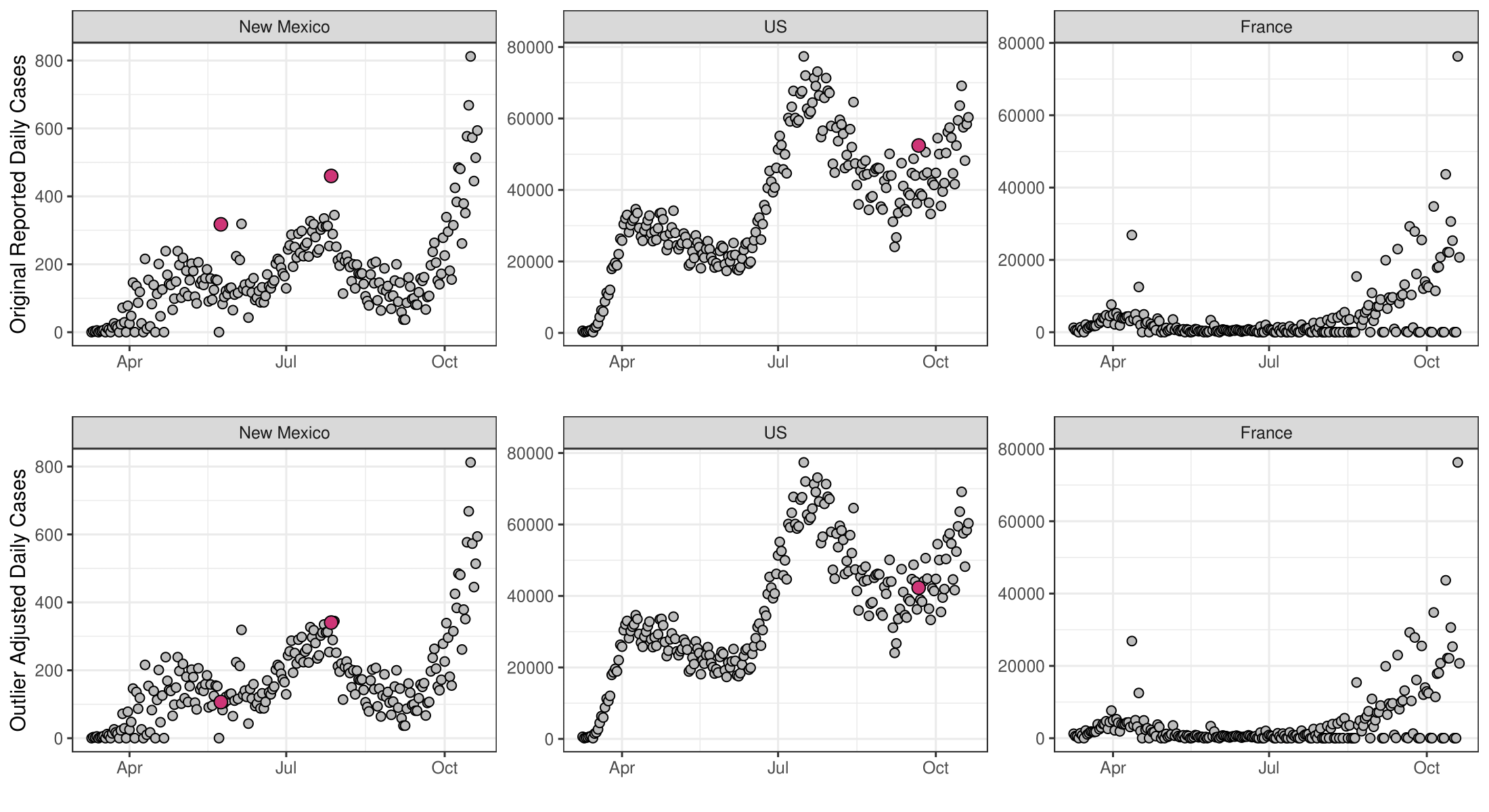}
          \caption{The reported daily cases of COVID-19 (top) and the outlier adjusted daily cases (bottom). (Top) Magenta points were identified as outliers. (Bottom) Magenta points are the adjusted outliers.}
          \label{fig:outlier_adjusted_cases}
\end{figure}

COFFEE automatically identifies and adjusts outliers \cite{tsoutliers}. It runs five different outlier detection algorithms on the reported data, taking into account possible day-of-week (DOW) effects. A datum is declared an outlier if three or more of the five detection algorithms identify that datum as an outlier. The outliers are not removed, but rather adjusted to ensure all values are non-negative. Figure \ref{fig:outlier_adjusted_cases} shows the result of this process on daily cases for New Mexico, the US, and France. All subsequent modeling steps are conducted with outlier adjusted data.

\subsubsection*{Step 2: Compute the Empirical Growth Rate, $\hat{\kappa}_t$}
The model for the underlying number of reported daily cases, $\delta_{c,t}$, is a dynamic susceptible-infectious (SI) model \cite{jacquez1993stochastic}, where 
\begin{align}
\label{eq:theta_s}
\delta_{s,t} &= \delta_{s,t-1} - \delta_{c,t}\\
\label{eq:theta_c}
\ddot{\delta}_{c,t} &= \ddot{\delta}_{c,t-1} + \delta_{c,t},
\end{align} 
\noindent and 
\begin{align}
\label{eq:delta_c}
\delta_{c,t} = \kappa_t \frac{\delta_{s,t-1}}{\delta_{s,0}} \ddot{\delta}_{c,t-1}.
\end{align}
\noindent  The quantity $\ddot{\delta}_{c,t-1}$ is the cumulative number of underlying cases on day $t-1$, $\frac{\delta_{s,t-1}}{\delta_{s,0}}$ is the proportion of the population still susceptible at time $t-1$, and $\kappa_t$ is the growth rate on day $t$.

\begin{figure}[h!]
    \centering
	 \includegraphics[width=1\linewidth]{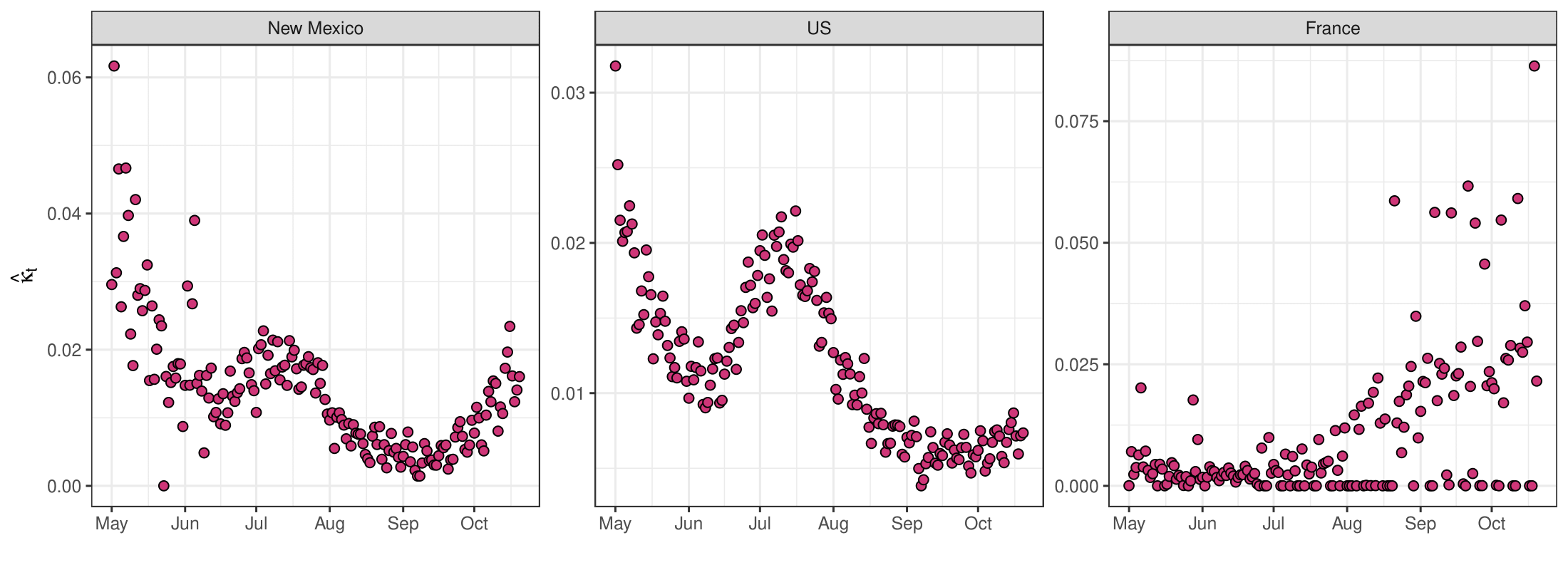}
          \caption{$\hat{\kappa}_t$ starting in May for New Mexico, the US, and France.}
          \label{fig:kappa_hat}
\end{figure}

When $\frac{\delta_{s,t-1}}{\delta_{s,0}} \approx 1$ (when most of the susceptible population is still susceptible), we can rearrange Equations \ref{eq:theta_s}, \ref{eq:theta_c}, and \ref{eq:delta_c} to identify a crude estimator of $\kappa_t$:
\begin{align}
\hat{\kappa}_t &\approx \Bigg(\frac{\ddot{y}_{c,t}}{\ddot{y}_{c,t-1}} - 1\Bigg).
\end{align}
\noindent Estimates for $\kappa_t$ are shown in Figure \ref{fig:kappa_hat}. It is clear that $\hat{\kappa}_t$ is dynamic and changes over time. This is what makes forecasting COVID-19 so challenging; parameters of epidemiologically-motivated models are dynamic and \emph{forecasting with them requires anticipating how these dynamic parameters will change in the future, not just tracking where they have been in the past}. In what follows, we describe how we forecast $\hat{\kappa}_t$.

\newpage 
\subsubsection*{Step 3: Compute $\hat{\kappa}^*_t$ = logit$(\hat{\kappa}_t)$}
\begin{figure}[h!]
    \centering
	 \includegraphics[width=1\linewidth]{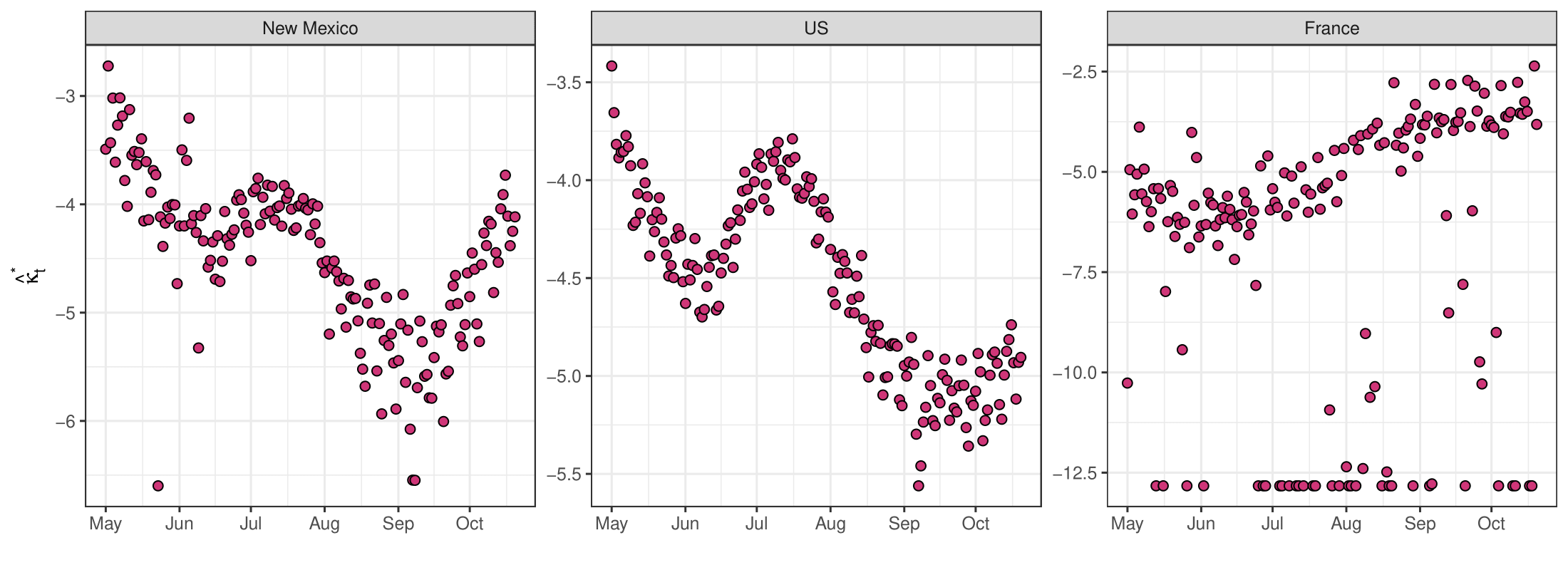}
          \caption{The quantities $\hat{\kappa}^*_t$ starting in May for New Mexico, the US, and France.}
          \label{fig:logit_kappa_hat}
\end{figure}

After the initial portion of the outbreak, $\hat{\kappa}_t$ is almost always between 0 and 1. COFFEE logit transforms $\hat{\kappa}_t$, where logit($p$) = log($p/(1-p)$) for $p \in (0,1)$. For all days with no reported cases, $\hat{\kappa}_t = 0$, an incompatible value with the logit transform. Thus, we compute $\hat{\kappa}^*_t = $ logit($\hat{\kappa}_t$) as follows:
\begin{align}
\hat{\kappa}^*_t &= 
\begin{cases}
\text{logit}(\hat{\kappa}_t)  & \text{ if } \hat{\kappa}_t > \tau_c \text{ and } \hat{\kappa}_t < 1 - \tau_c\\
\text{logit}(\tau_c)  & \text{ if } \hat{\kappa}_t \leq \tau_c\\
\text{logit}(1-\tau_c)  & \text{ if } \hat{\kappa}_t \geq 1-\tau_c,\\
\end{cases}
\end{align}
\noindent where $\tau_c = 0.95*\text{min}(\{\hat{\kappa}_t | \hat{\kappa}_t > 0\})$, ensuring that if $\hat{\kappa}_t \geq \hat{\kappa}_{t'}$, then $\hat{\kappa}^*_t \geq \hat{\kappa}^*_{t'}$. The logit transformed $\hat{\kappa}_t$ are shown in Figure \ref{fig:logit_kappa_hat}.

\subsubsection*{Step 4: Split Data into Training and Testing Sets}
\begin{figure}[h!]
    \centering
	 \includegraphics[width=1\linewidth]{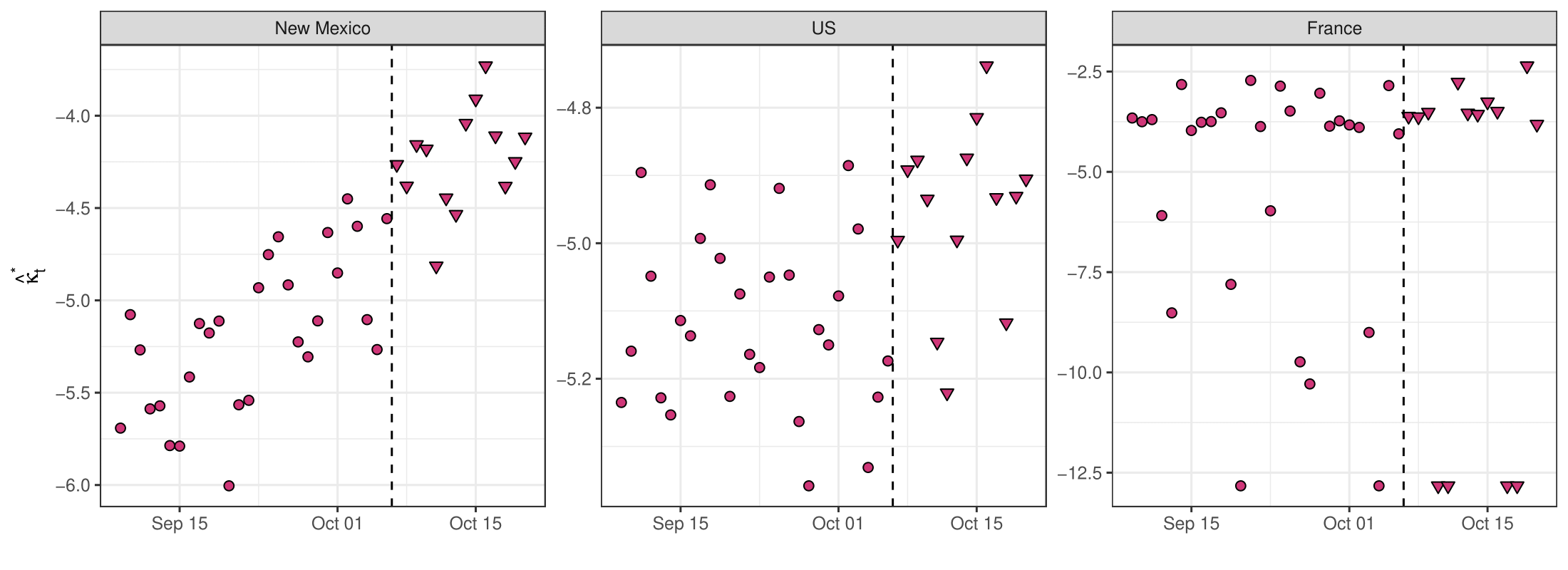}
          \caption{The quantities $\hat{\kappa}^*_t$ for New Mexico, the US, and France. Circles are training data, left of the vertical dashed line, while testing data are the triangles to the right of the dashed vertical line.}
          \label{fig:splits}
\end{figure}

Let $T$ be the last observed day. We only consider the last 42 days of data when fitting a model for $\hat{\kappa}^*_t$ and split those days into training and testing data \cite{picard1990data}. Days $T-41$ through $T-14$ constitute the training data, while $T-13$ through $T$ constitutes the testing data. We will denote the last day of the training data by $T^{\text{train}} = T-14$. The splits are shown for New Mexico, the US, and France in Figure \ref{fig:splits}.

\subsubsection*{Step 5: Compute $\hat{\kappa}_t^{\text{trend}}$}
\begin{figure}[h!]
    \centering
	 \includegraphics[width=1\linewidth]{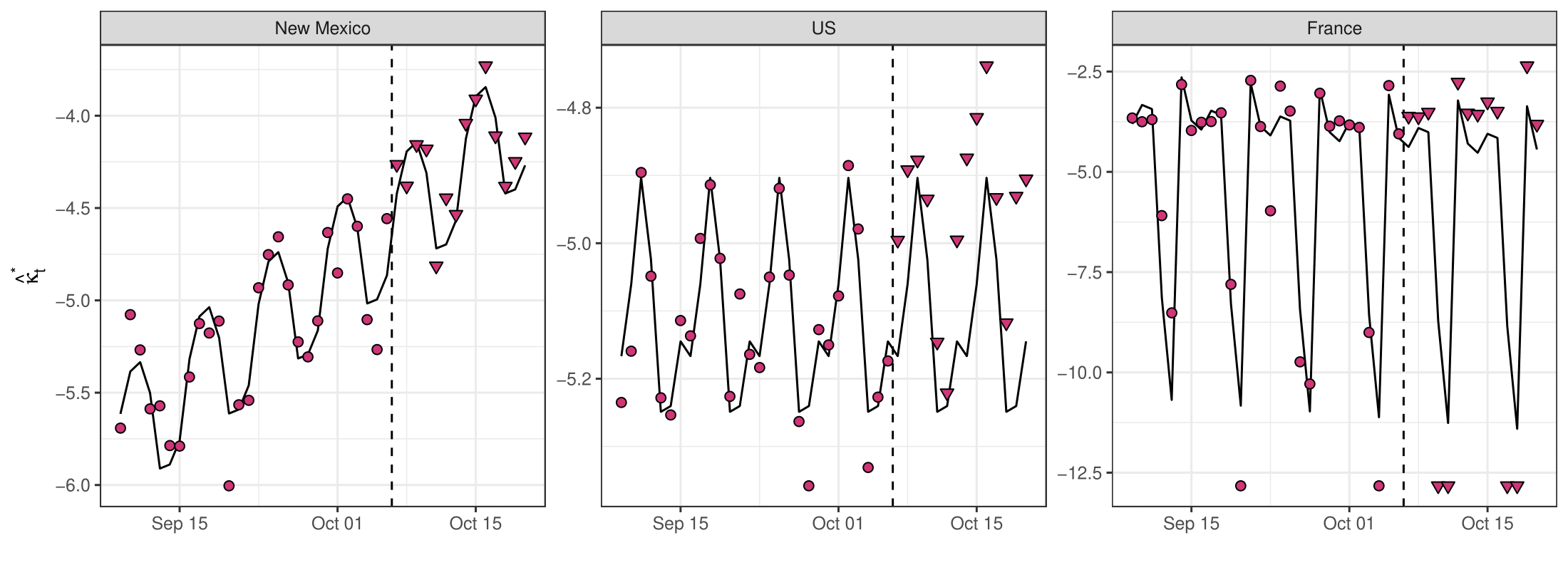}
          \caption{The quantities $\hat{\kappa}^*_t$ for New Mexico, the US, and France. Circles are training data, left of the vertical dashed line, while testing data are the triangles to the right of the dashed vertical line. Solid line represents the fits (train) and predictions (test) of $\hat{\kappa}_t^{\text{trend}}$ based on the regression.}
          \label{fig:kappa_reg}
\end{figure}

We fit a weighted regression to the training data, where we downweight influential points that could have an outsized influence on the regression using the inverse of Cook's distance. The regression has a linear trend over time and a DOW effect:
\begin{align}
\label{eq:reg}
\hat{\kappa}^*_t &= \beta_0 + \beta_1 t + \beta_2 \text{I}(t = \text{Monday}) +  \beta_3 \text{I}(t = \text{Tuesday}) + \ldots + \beta_7 \text{I}(t = \text{Saturday}).
\end{align}
\noindent Variable selection is performed, potentially resulting in a subset of the model parameters in Equation \ref{eq:reg}. We refer to the fits and predictions from this linear model as $\hat{\kappa}_t^{\text{trend}}$ where
\begin{align}
\label{eq:dow}
\hat{\kappa}_t^{\text{trend}} = \hat{\beta}_0 + \hat{\beta}_1 t + \hat{\beta}_2 \text{I}(t = \text{Monday}) +  \hat{\beta}_3 \text{I}(t = \text{Tuesday}) + \ldots + \hat{\beta}_7 \text{I}(t = \text{Saturday})
\end{align}
\noindent where, if a variable was removed during the variable selection phase, then the corresponding $\hat{\beta}$ is set equal to 0. Figure \ref{fig:kappa_reg} shows $\hat{\kappa}^{\text{trend}}_t$ for the training and testing windows.

\subsubsection*{Step 6: Compute $\hat{\kappa}_t^{\text{constant}}$}
There is a $\hat{\kappa}_t$ trajectory corresponding to a constant new number of cases day over day. Let
\begin{align}
\label{eq:avg_cases}
\bar{y}_{c,T^{\text{train}}} &= \frac{1}{7}\sum_{t=T^{\text{train}}-6}^{T^{\text{train}}} y_{c,t}
\end{align}
\noindent be the average number of daily reported cases over the last week of the training window.

\begin{figure}[h!]
    \centering
	 \includegraphics[width=1\linewidth]{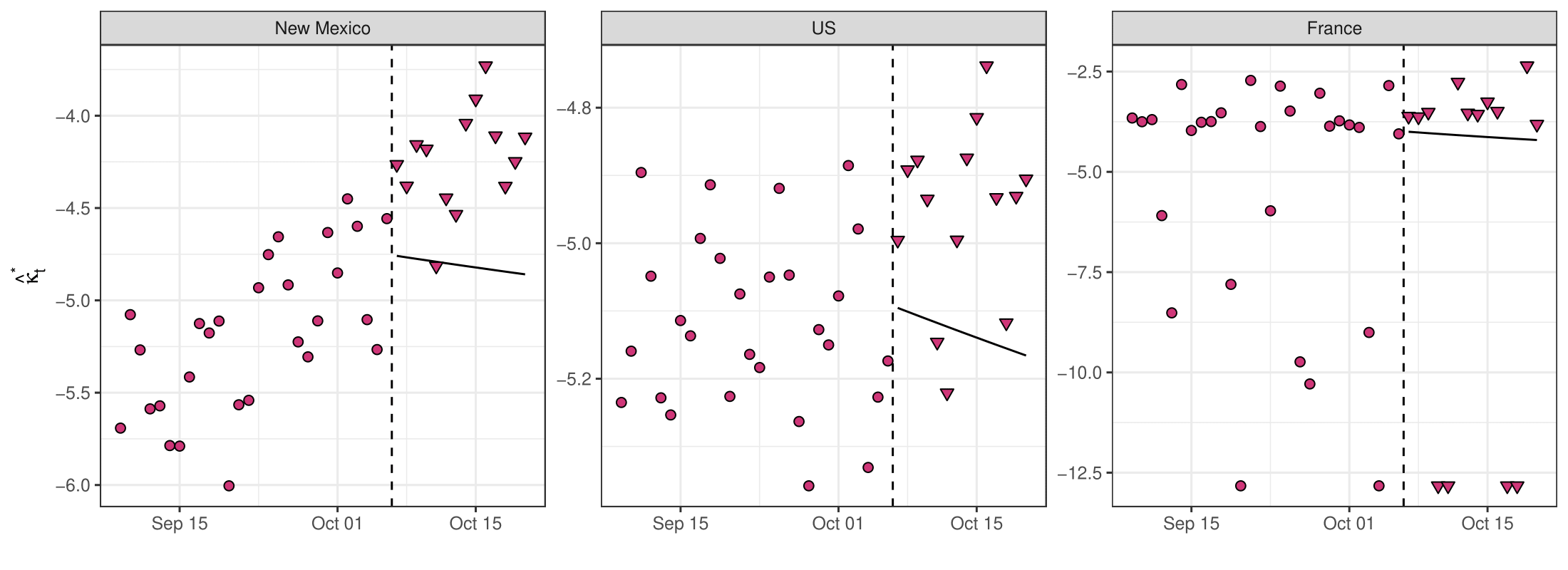}
          \caption{The quantities $\hat{\kappa}^*_t$ for New Mexico, the US, and France. Circles are training data, left of the vertical dashed line, while testing data are the triangles to the right of the dashed vertical line. Solid line represents the values of $\hat{\kappa}_t^{\text{constant}}$, the trajectory corresponding to a constant number of new reported cases, equal to $\bar{y}_{c,T^{\text{train}}}$.}
          \label{fig:kappa_constant}
\end{figure}

Then 
\begin{align}
\label{eq:kappa_constant}
\hat{\kappa}_{t}^{\text{constant}} &= \text{logit}\Bigg(\bar{y}_{c,T^{\text{train}}} \Bigg[ \Bigg( \frac{\delta_{s,0} - \ddot{y}_{c,t-1} }{\delta_{s,0}} \Bigg) \ddot{y}_{c,t-1} \Bigg]^{-1}\Bigg)
\end{align}
\noindent where we set $\delta_{s,0} = 0.55 N$, where $N$  is the population of the forecasted region and 0.55 is a nominal attack rate for COVID-19. Figure \ref{fig:kappa_constant} shows $\hat{\kappa}_{t}^{\text{constant}}$ for $t = T^{\text{train}} + k$ and $k \in 1,2,\ldots, 14$.

\subsubsection*{Step 7: Compute a Joint Probability Distribution over Tuning Parameters}
The form of the forecasting model for $\hat{\kappa}_t^{\text{forecast}}$ is
\begin{align}
\label{eq:fcstform}
\hat{\kappa}_t^{\text{forecast}}(\eta,\omega,\phi) &=  \lambda_t(\phi) [w_t \text{min}(\eta^*,\hat{\kappa}_t^{\text{trend}}) + (1-w_t)\hat{\kappa}_t^{\text{constant+DOW}}]
\end{align}
\noindent for $t = T^{\text{train}} + k$ and $k \in 1,2,\ldots, 14$. The objective is to produce forecasts for $\hat{\kappa}_t$. The way COFFEE does this is be creating a blended combination of $\hat{\kappa}_t^{\text{trend}}$ and $\hat{\kappa}_t^{\text{constant+DOW}}$ where 
\begin{align}
\hat{\kappa}_t^{\text{constant+DOW}} =& \hat{\kappa}_t^{\text{constant}} +  \nonumber \\
&\hat{\beta}_2 \text{I}(t = \text{Monday}) +  \hat{\beta}_3 \text{I}(t = \text{Tuesday}) + \ldots + \hat{\beta}_7 \text{I}(t = \text{Saturday}),
\end{align} 
\noindent which is $\hat{\kappa}_t^{\text{constant}}$ with the DOW effects estimated in Equation \ref{eq:dow} added.

There are three tuning parameters in Equation \ref{eq:fcstform}, each playing a role in controlling the form of $\hat{\kappa}_t^{\text{forecast}}$.

The first tuning parameter $\eta$ puts a cap on how large $\hat{\kappa}_t^{\text{trend}}$ can get. A forecast can blow up if $\hat{\kappa}_t^{\text{trend}}$ is growing in an unmitigated fashion. The parameter $\eta$ is a safeguard against this unmitigated growth. We set
\begin{align}
\eta^* &= \text{median}(\{\hat{\kappa}^*_{T^{\text{train}}-6}, \hat{\kappa}^*_{T^{\text{train}}-5},\ldots,\hat{\kappa}^*_{T^{\text{train}}}\})\eta
\end{align}
\noindent for $\eta \in [0,1]$.

\begin{figure}[h!]
    \centering
	 \includegraphics[width=1\linewidth]{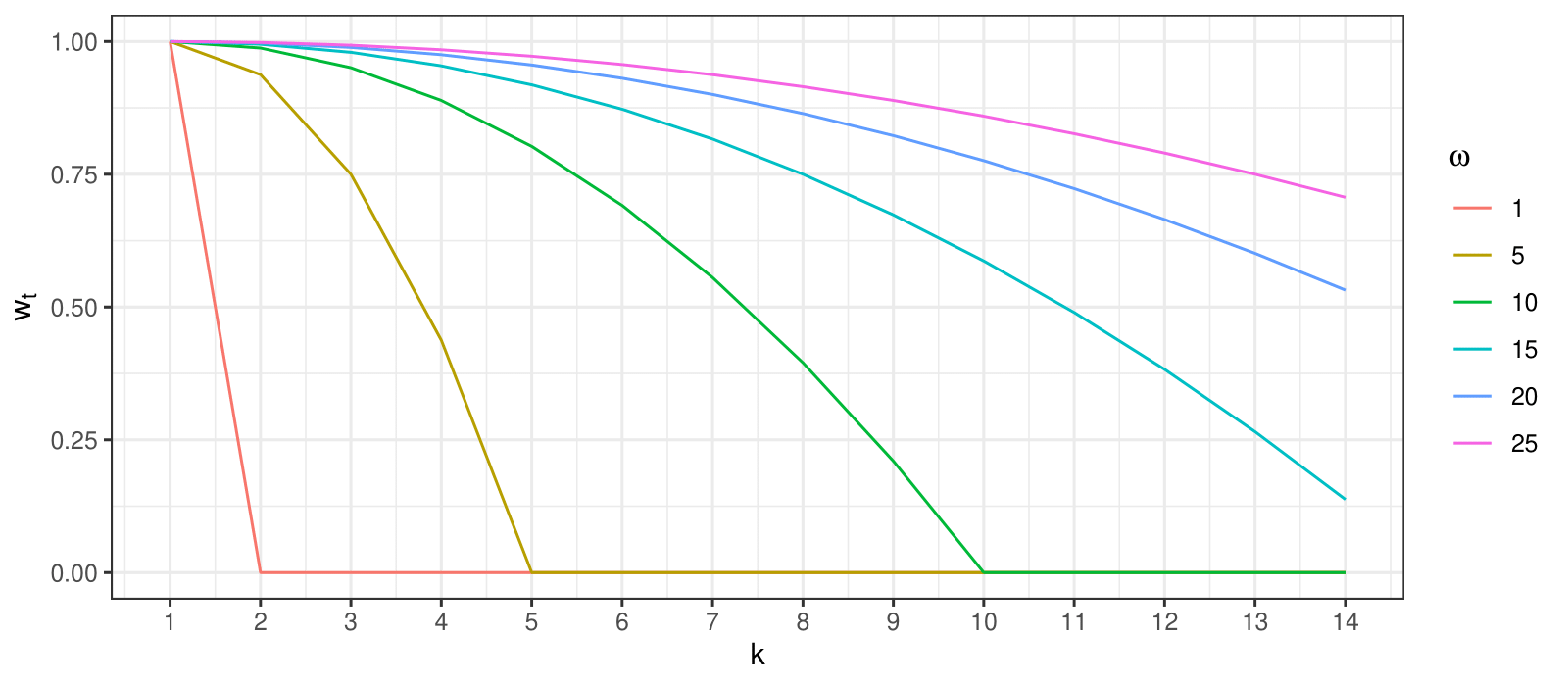}
          \caption{Weight trajectories $w_{T^{\text{train}}+k}$ for different values of $\omega$. The smaller $\omega$ is, the quicker $w_{T^{\text{train}}+k}$ transitions from 1 where all weight is assigned to $\hat{\kappa}^{\text{trend}}$ to 0 where all weight is assigned to $\hat{\kappa}_t^{\text{constant+DOW}}$.}
          \label{fig:omega}
\end{figure}

The second tuning parameter is $\omega$. The basic form of the forecasting model is to transition from a forecast that relies on the current trend $\hat{\kappa}^{\text{trend}}_t$ to a forecast that relies on $\hat{\kappa}^{\text{constant+DOW}}_t$. If $\hat{\kappa}_t^{\text{trend}}$ is trending up, this transition keeps the forecasts from blowing up. If $\hat{\kappa}_t^{\text{trend}}$ is trending down, this transition keeps the forecasts from flat-lining at 0 new cases. The assumption behind this modeling choice is that, as cases are going up, people will take action to curb the growth of the pandemic, either through independent choices of personal responsibility or governmental policies. As cases are going down, however, we assume policies will be relaxed or people will become more comfortable engaging in activities that will increase transmission pathways. The tuning parameter $\omega \geq 1$ determines how quickly the forecast transitions from $\hat{\kappa}_t^{\text{trend}}$ to $\hat{\kappa}_t^{\text{constant+DOW}}$. The closer $\omega$ is to 0, the quicker the transition occurs.
\begin{align}
w_{T^{\text{train}}+k} &= 
\begin{cases}
1-\Bigg(\frac{k-1}{\omega} \Bigg)^2 & \text{ if } k \leq \omega + 1\\
0 & \text{ otherwise }
\end{cases}
\end{align}
\noindent where $k$ is a positive integer. Figure \ref{fig:omega} shows weight trajectories $w_t$ for various choices of $\omega$. When $w_{T^{\text{train}}+k} = 1$, all weight is on $\hat{\kappa}_t^{\text{trend}}$; when $w_{T^{\text{train}}+k}=0$, all weight is on $\hat{\kappa}_t^{\text{constant+DOW}}$.

\begin{figure}[h!]
    \centering
	 \includegraphics[width=1\linewidth]{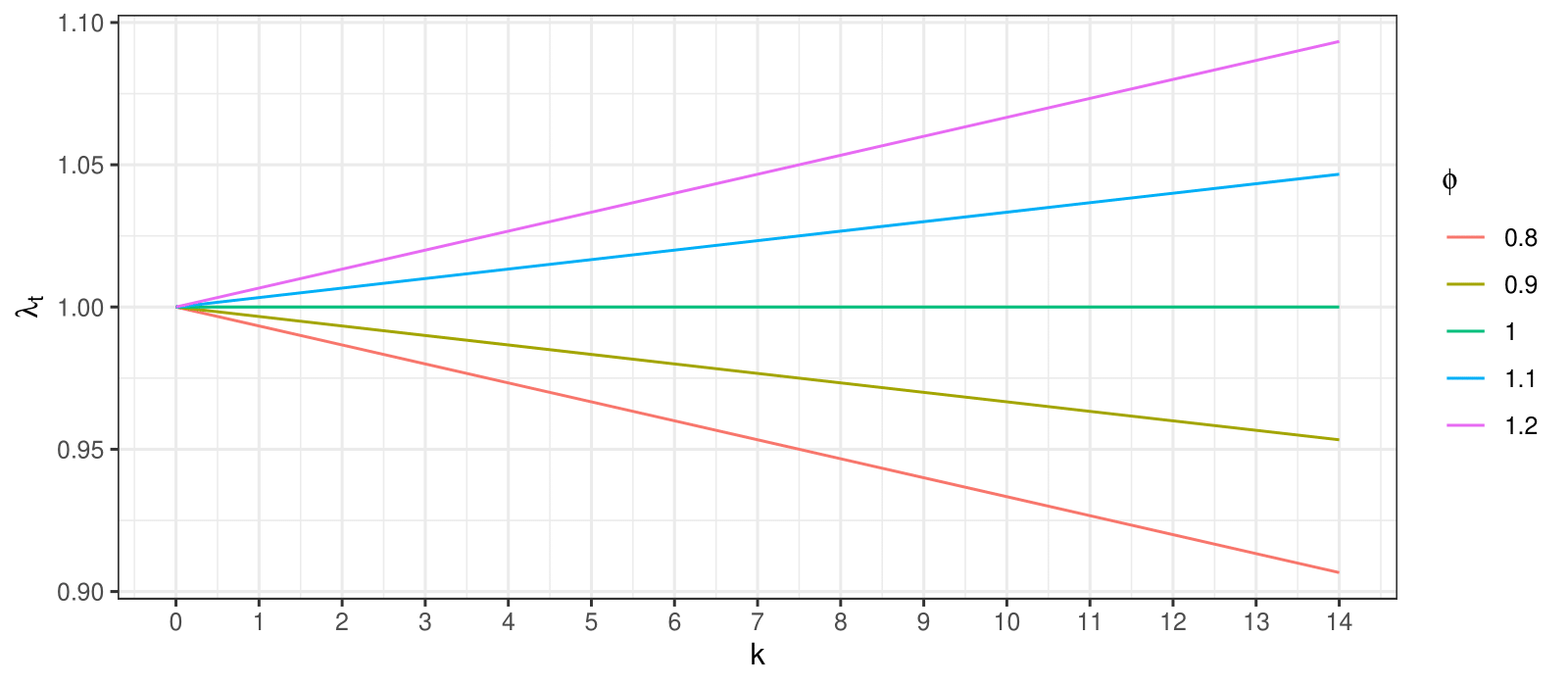}
          \caption{The trajectories $\lambda_{T^{\text{train}}+k}$ for different values of $\phi$. For $\phi$ less than 1, $\lambda_{T^{\text{train}}+k}$ tilts the blended combination of $\hat{\kappa}^{\text{trend}}_t$ and $\hat{\kappa}^{\text{constant+DOW}}_t$ up. When $\phi$ is greater than 1, $\lambda_{T^{\text{train}}+k}$ tilts it down.}
          \label{fig:lambda}
\end{figure}

The third tuning parameter is $\phi$. The trajectory $\lambda_t$ is defined as
\begin{align}
\lambda_{T^{\text{train}}+k} &= 1 + k\frac{\phi - 1}{30},
\end{align}
\noindent a linear trend starting at 1 when $k=0$. The tuning parameter $\phi > 0$ determines whether $\lambda_{T^{\text{train}}+k}$ trends up ($\phi > 1$) or down ($\phi < 1$). Examples of $\lambda_{T^{\text{train}}+k}$ are shown in Figure \ref{fig:lambda}.

For a combination of $\eta$, $\omega$, and $\phi$, we compute $\hat{\kappa}_t^{\text{forecast}}(\eta,\omega,\phi)$ and compute the inverse-distance between the inverse-logit of $\hat{\kappa}_t^{\text{forecast}}(\eta,\omega,\phi)$ and $\hat{\kappa}_t$ over the test period:
\begin{align}
\label{eq:invdist}
d^{-1}(\eta,\omega,\phi) &= \Bigg(\sum_{t=T^{\text{train}}+1}^{T^{\text{train}}+14} \Bigg[ \text{logit}^{-1}\Bigg(\hat{\kappa}_t^{\text{forecast}}(\eta,\omega,\phi)\Bigg) - \hat{\kappa}_t \Bigg]^2\Bigg)^{-1}.
\end{align}

\begin{figure}[h!]
    \centering
	 \includegraphics[width=1\linewidth]{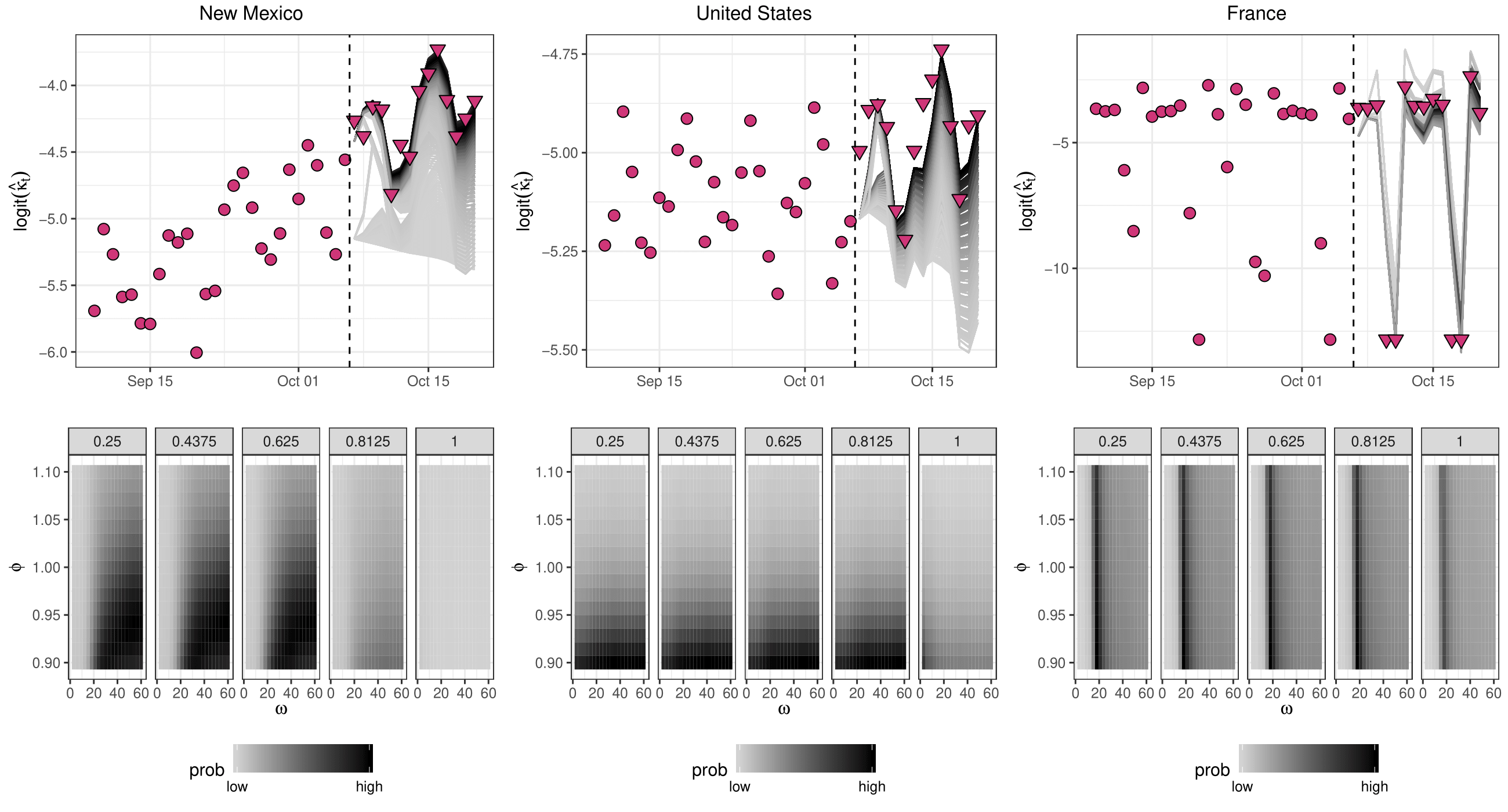}
          \caption{(Top) $\hat{\kappa}^*_t$ for the training period (circles) and testing period (triangles). Lines in the testing period are $\hat{\kappa}_t^{\text{forecast}}$. Each line corresponds to a combination of $\eta$, $\omega$, and $\phi$. The color of the line is proportional to $d^{-1}(\eta,\omega,\phi)$ with darker lines corresponding to better agreement between $\hat{\kappa}^{\text{forecast}}_t$ and $\hat{\kappa}^*_t$ in the test set. (Bottom) The normalized inverse-distance values for $\omega$ (x-axis), $\phi$ (y-axis), and $\eta$ (panels). Darker tiles correspond to larger inverse-distances.}
          \label{fig:train}
\end{figure}

Finally we compute a joint probability distribution over $\eta$, $\omega$, and $\phi$ as the normalized inverse-distance. The probability distributions for New Mexico, the US, and France are shown in Figure \ref{fig:train}. 

\subsubsection*{Step 8: Produce the Reported Cases Forecast}
The final step is to simulate reported cases. The purpose of the previous steps was to get a joint probability distribution over the tuning parameters that can be used to sample from. 

If more than 14 of the last 28 days had zero reported cases, we take independent and identically distributed (iid) samples of future reported cases from the empirical distribution of outlier adjusted reported cases over the last 28 days. 
If no cases were reported over the last 28 days, we sample future reported cases as iid Bernoulli draws with success probability equal to 1/29. 
If 14 or more of the last 28 days observed at least 1 reported case, we simulate $\hat{\kappa}^{\text{forecast}}_{T+k}$ for $k \in {1,2,...,K}$ by doing the following:

\begin{enumerate}
\item Do Step 5, treating the training data as days $T$ to $T - 27$, resulting in a fitted linear model in the form of Equation \ref{eq:reg}. Use this to compute $\hat{\kappa}^{\text{trend}}_{T+k}$.
\item Do Step 6 to compute $\hat{\kappa}^{\text{constant+DOW}}_{T+k}$, replacing $T^{\text{train}}$ with $T$ in Equations \ref{eq:avg_cases} and \ref{eq:kappa_constant}.
\item Draw a vector of $(\omega, \phi, \eta)$ from the joint distribution computed in Step 7. 
\item Compute $\hat{\kappa}_{T+k}^{\text{forecast}}$ following Equation \ref{eq:fcstform}.
\item Compute logit$^{-1}(\hat{\kappa}_{T+k}^{\text{forecast}})$.
\item Draw an attack rate $p \sim \text{Uniform}(0.4,0.7)$ and set $\delta^{\text{forecast}}_{s,0} = pN$.
\item Set $\ddot{\delta}_{c,T} = \ddot{y}_{c,T}$ and $\delta_{s,T} = \delta^{\text{forecast}}_{s,0} - \ddot{y}_{c,T}$
\item For $k=1,2,\ldots,K$, compute
\begin{enumerate}
\item $\delta^{\text{forecast}}_{c,T+k} = \text{logit}^{-1}(\hat{\kappa}_{T+k}^{\text{forecast}})\Bigg(\frac{\delta^{\text{forecast}}_{s,T+k-1}}{\delta^{\text{forecast}}_{s,0}} \Bigg) \ddot{\delta}^{\text{forecast}}_{c,T+k-1}$
\item $\ddot{\delta}^{\text{forecast}}_{c,T+k} = \ddot{\delta}^{\text{forecast}}_{c,T+k-1} + \delta^{\text{forecast}}_{c,T+k}$
\item $\delta^{\text{forecast}}_{s,T+k} = \delta^{\text{forecast}}_{s,T+k-1} - \delta^{\text{forecast}}_{c,T+k}$
\end{enumerate}
\item Draw $y^{\text{forecast}}_{c,t}|\delta^{\text{forecast}}_{c,T+k},\hat{\alpha} \sim \text{NB}\Bigg(\delta^{\text{forecast}}_{c,T+k}, \frac{\delta^{\text{forecast}}_{c,T+k}}{\hat{\alpha}}\Bigg)$ where $\hat{\alpha}$ is the maximum likelihood estimate.
\end{enumerate}

\begin{figure}[h!]
   \centering
	 \includegraphics[width=1\linewidth]{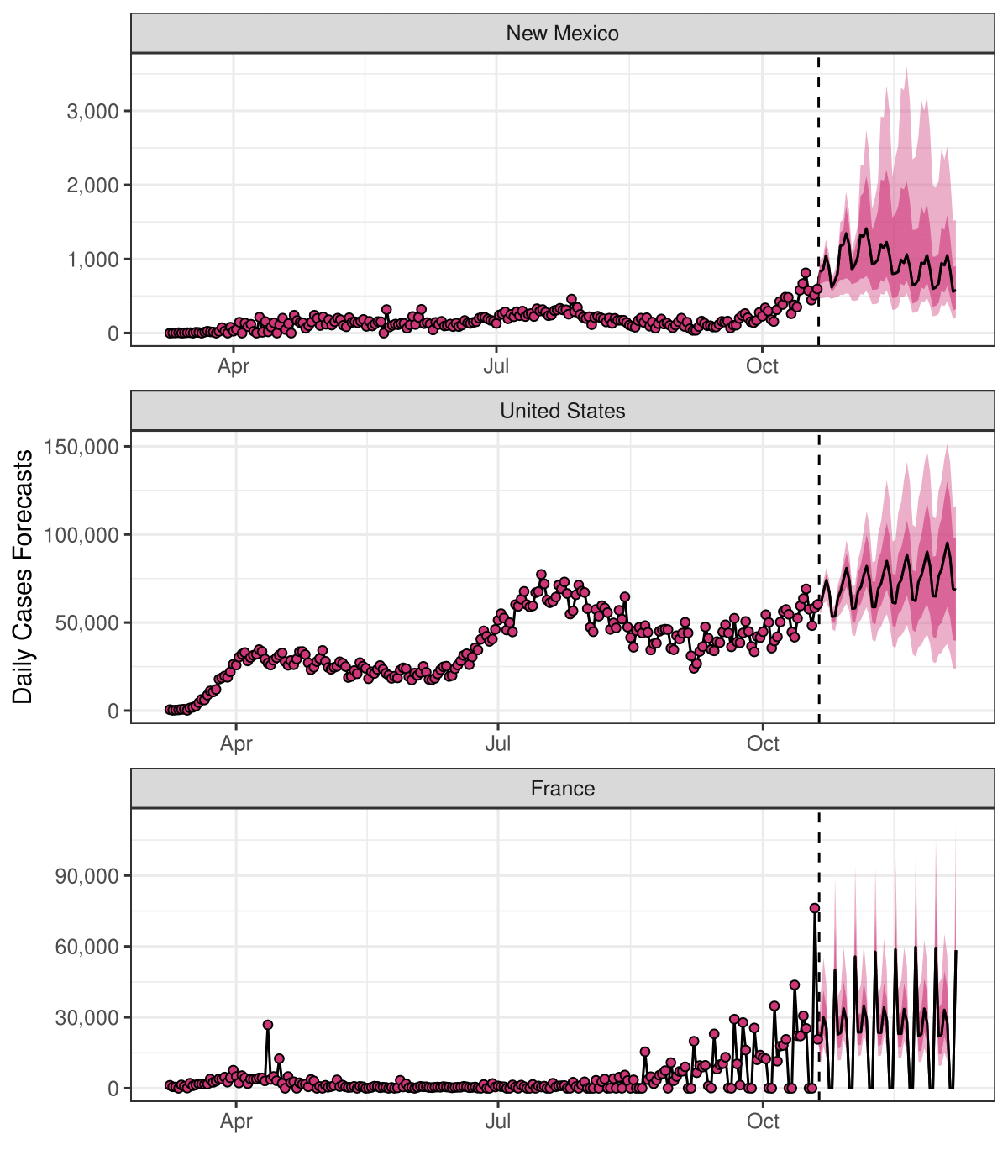}
        \caption{The median (black line) and 50\% and 80\% prediction intervals (ribbons) for New Mexico, the US, and France for daily reported cases.}
      \label{fig:cases_fcsts}
\end{figure}

Figure \ref{fig:cases_fcsts} shows the forecasts for New Mexico, the US, and France.

\newpage

\subsection*{Deaths Model} 
\begin{figure}[h!]
   \centering
	 \includegraphics[width=1\linewidth]{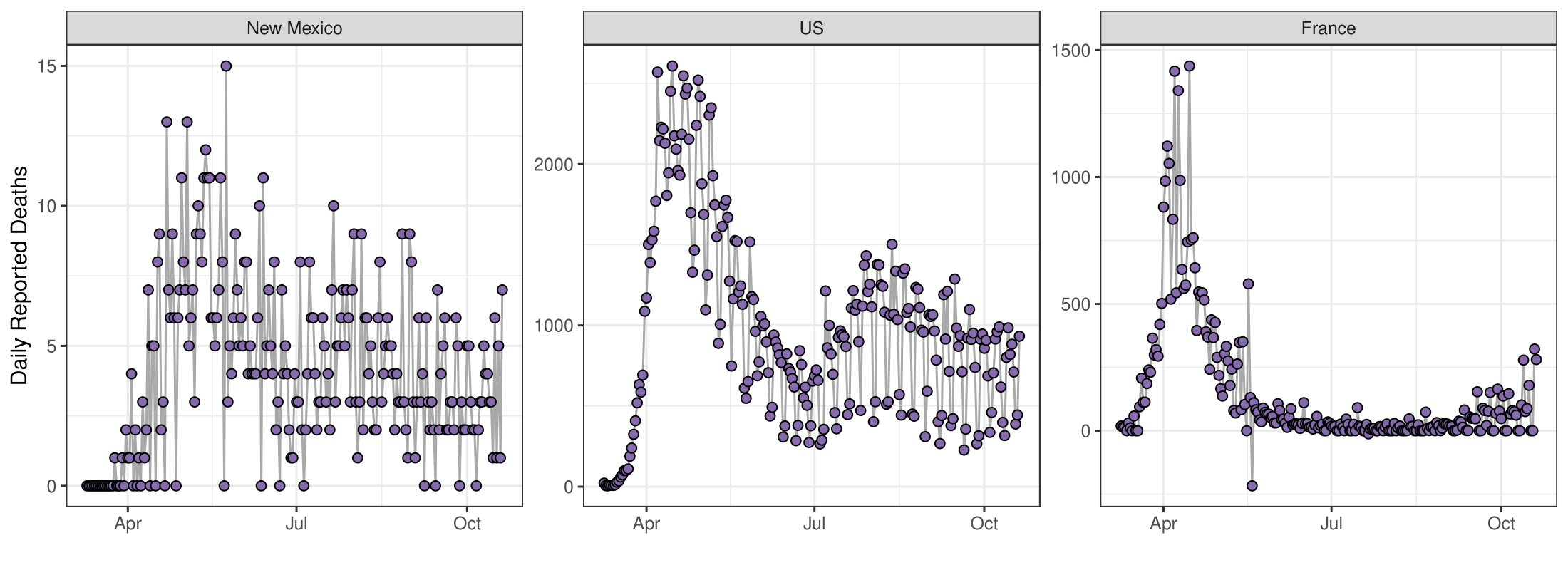}
        \caption{Daily reported deaths for New Mexico, the US, and France.}
      \label{fig:deaths}
\end{figure}

Figure \ref{fig:deaths} shows the daily deaths for New Mexico, the US, and France. 

The COFFEE deaths model is
\begin{align}
\label{eq:gamma}
\delta_{d,t} &= \gamma_t f(\boldsymbol{\delta}_{c,1:t},\nu)
\end{align}
\noindent where $\gamma_t$ is the case fatality ratio and $f(\boldsymbol{\delta}_{c,1:t},\nu)$ is a moving average of $\boldsymbol{\delta}_{c,1:t}$ with window size equal to $\nu$:

\begin{align}
f(\boldsymbol{\delta}_{c,1:t},\nu) &= \frac{1}{\nu}\sum_{j=t-\nu+1}^t \delta_{c,j}.
\end{align}

\noindent The deaths model proceeds with the following steps.

\subsubsection*{Step 1: Identify and Adjust Outliers}
\begin{figure}[h!]
    \centering
	 \includegraphics[width=1\linewidth]{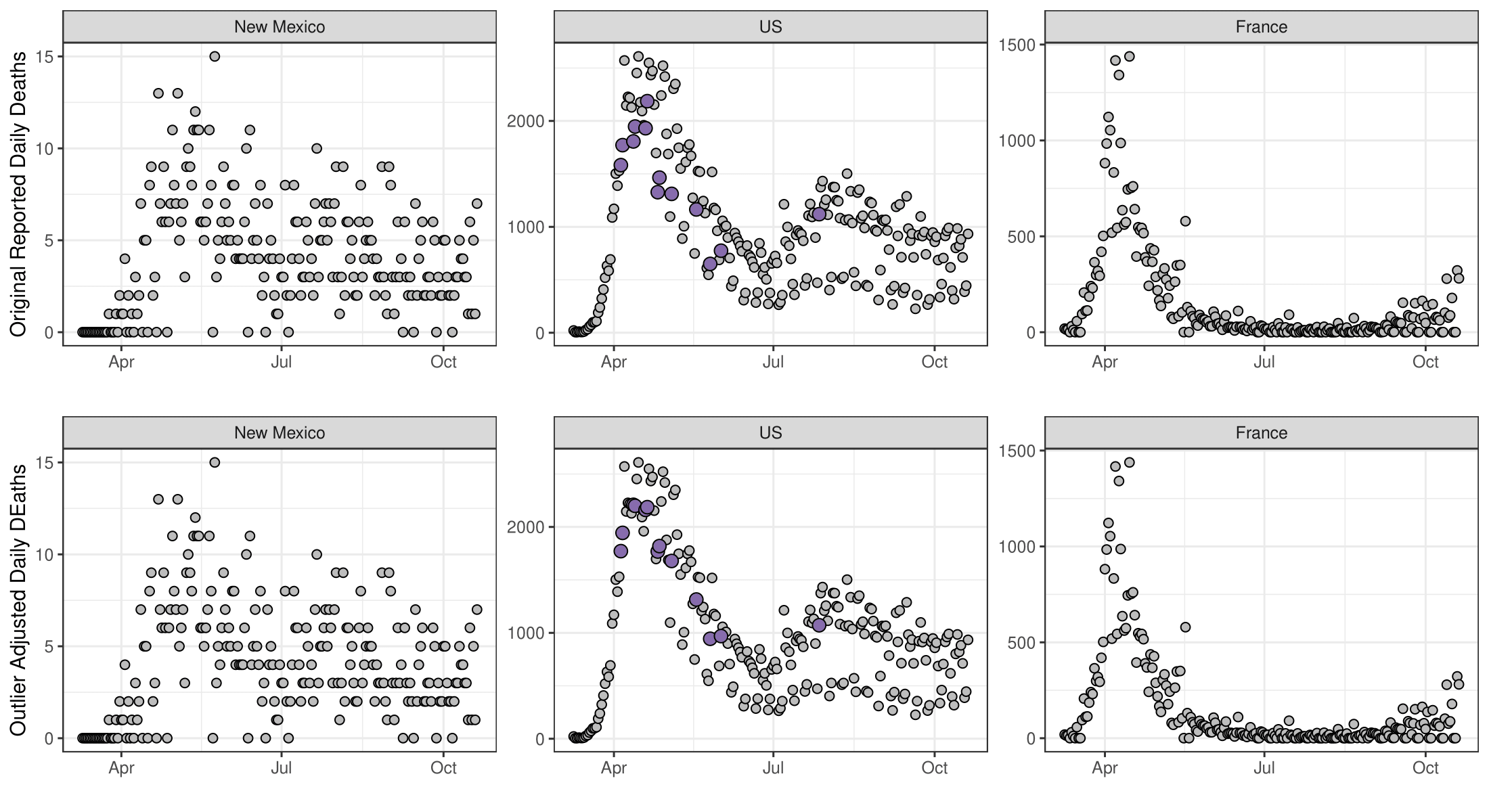}
          \caption{The originally reported daily deaths of COVID-19 (top) and the outlier adjusted daily deaths (bottom). (Top) Purple points were identified as outliers. (Bottom) Purple points are the adjusted outliers.}
          \label{fig:deaths_outliers}
\end{figure}

COFFEE uses the same outlier identification and adjustment routine as with cases, resulting in outlier adjusted deaths which are used for all subsequent forecasting steps. The outlier adjusted deaths are shown in Figure \ref{fig:deaths_outliers}.

\subsubsection*{Step 2: Compute the Case Fatality Ratio, $\hat{\gamma}_t$}
\begin{figure}[h!]
    \centering
	 \includegraphics[width=1\linewidth]{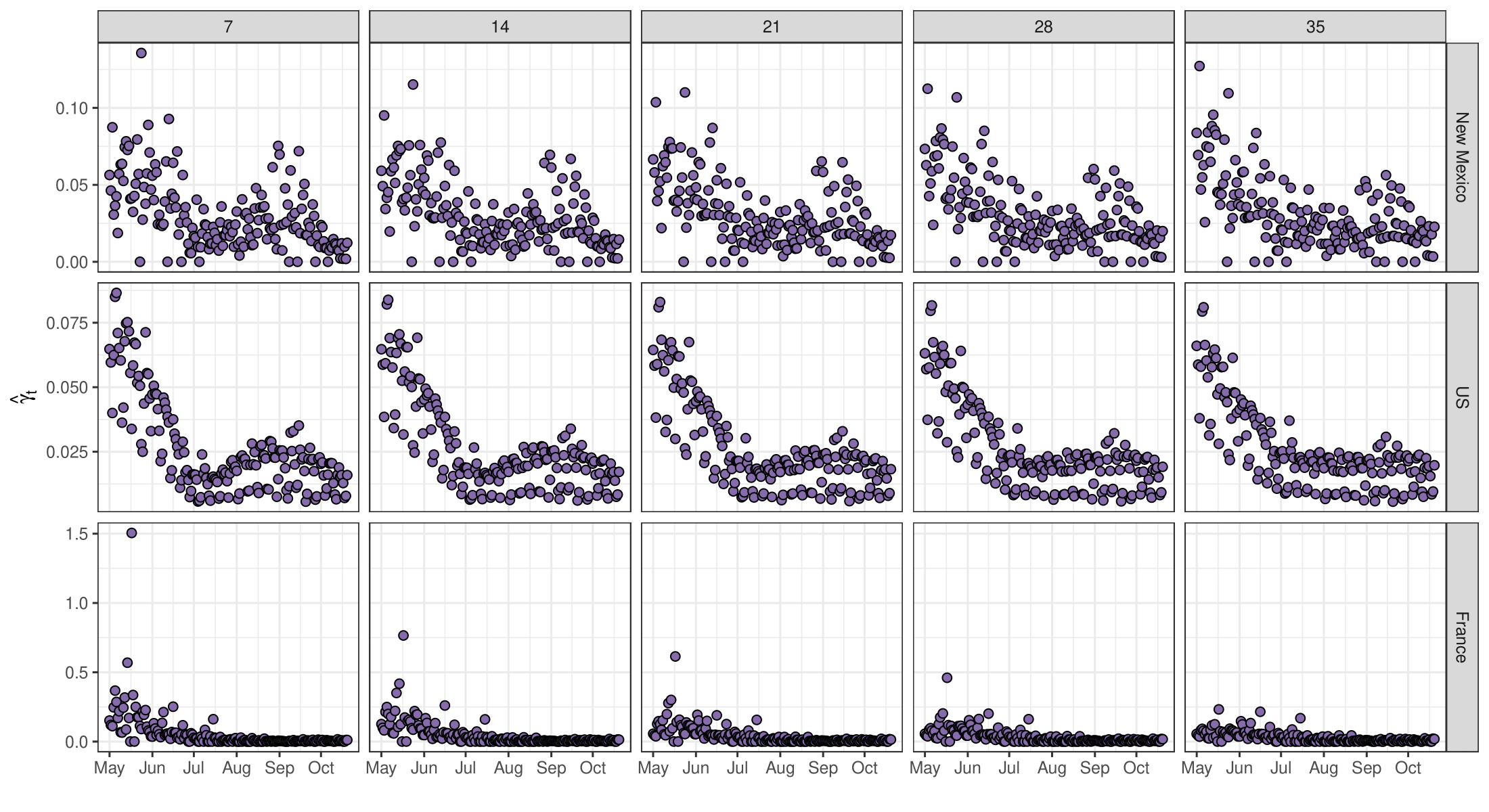}
          \caption{The values $\hat{\gamma}_t$ for New Mexico, the US, and France (rows) for different moving average window sizes of $\nu$ (columns).}
          \label{fig:gamma_hat}
\end{figure}

We estimate $\gamma_t$ by rearranging Equation \ref{eq:gamma} and replacing $\delta_{d,t}$ with $y_{d,t}$ and $\boldsymbol{\delta}_{c,1:t}$ with $\boldsymbol{y}_{c,1:t}$ for $t \leq T$:
\begin{align}
\hat{\gamma}_t &= y_{d,t}/f(\boldsymbol{y}_{c,1:t},\nu).
\end{align}
\noindent Figure \ref{fig:gamma_hat} displays $\hat{\gamma}_t$ for $\nu \in 7, 14, 21, 28, 35$.

\subsubsection*{Step 3: Compute $\hat{\gamma}^*_t = \text{logit}(\hat{\gamma}_t)$}
\begin{figure}[h!]
    \centering
	 \includegraphics[width=1\linewidth]{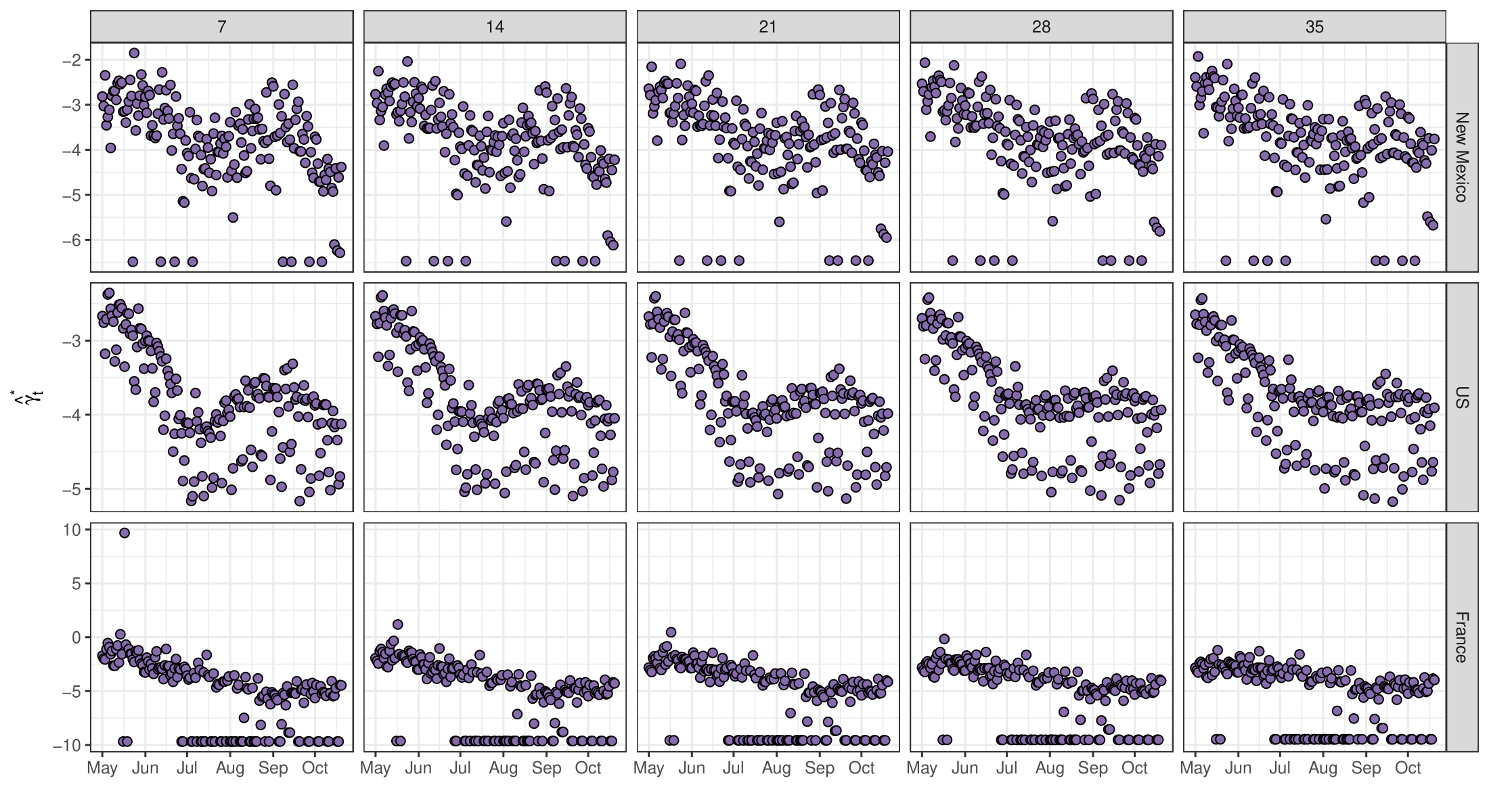}
          \caption{The values of $\hat{\gamma}^*_t$ for New Mexico, the US, and France (rows) for different values of $\nu$ (columns).}
          \label{fig:logit_gamma_hat}
\end{figure}

COFFEE logit transforms $\hat{\gamma}_t$, setting all values of $\hat{\gamma}_t < \tau_d$ equal to $\tau_d$ and all values of $\hat{\gamma}_t > 1-\tau_d$ equal to $1-\tau_d$ where $\tau_d = 0.95*\text{min}(\{\hat{\gamma}_t | \hat{\gamma}_t > 0\})$. The logit transformed $\hat{\gamma}_t$ are shown in Figure \ref{fig:logit_gamma_hat}.

\subsubsection*{Step 4: Split Data into Training and Testing Sets}
Split $\hat{\gamma}^*_t$ in a training and testing data set, same as with the cases model.

\subsubsection*{Step 5: Compute $\hat{\gamma}_t^{\text{trend}}$}
\begin{figure}[h!]
    \centering
	 \includegraphics[width=1\linewidth]{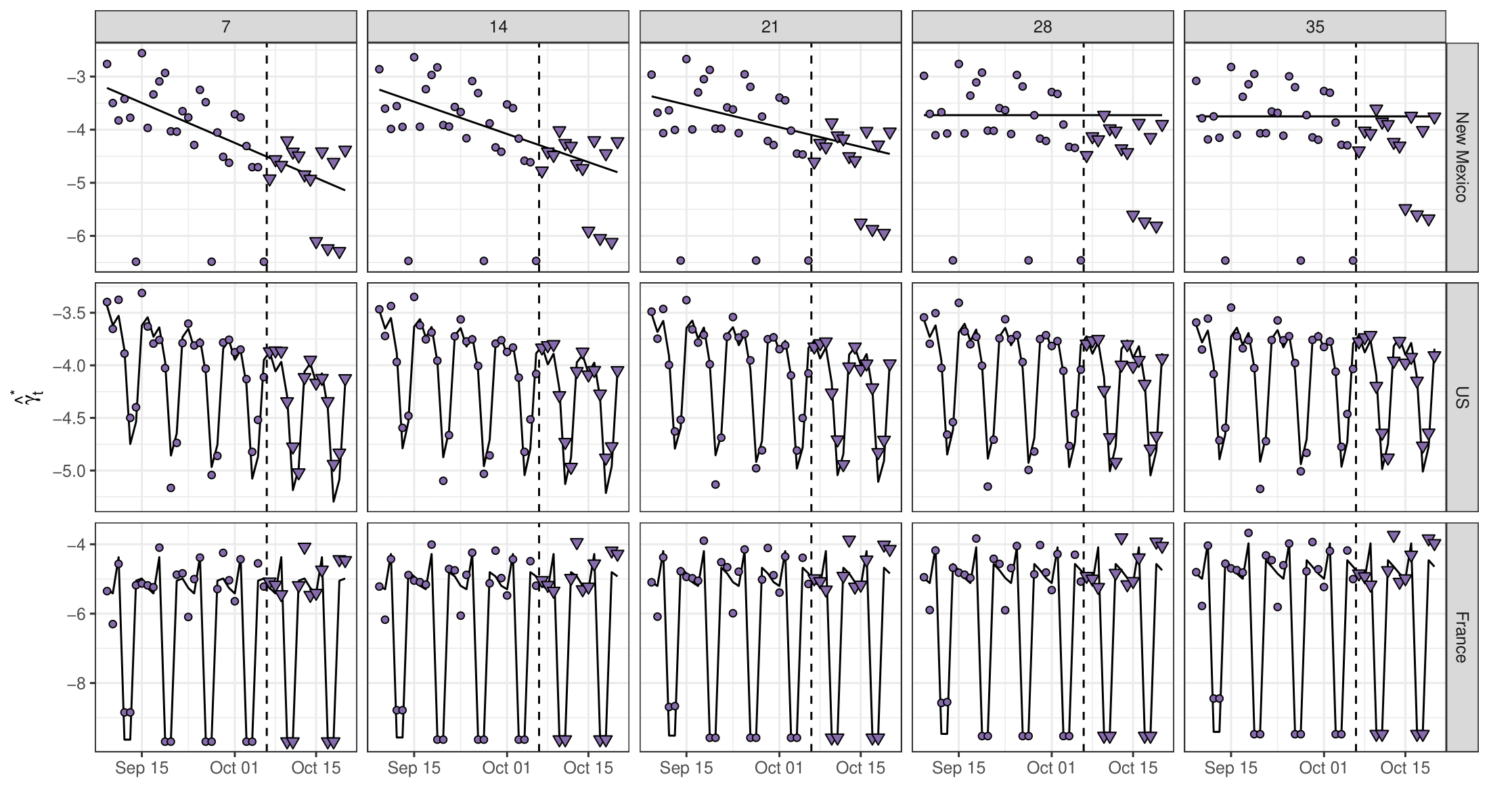}
          \caption{The quantities $\hat{\gamma}^*_t$ for New Mexico, the US, and France. Circles are training data, left of the vertical dashed line, while testing data are the triangles to the right of the dashed vertical line. Solid line represents the fits (train) and predictions (test) of $\hat{\gamma}_t^{\text{trend}}$ based on the regression. The US and France have a DOW effect, while New Mexico had the DOW effect removed in the variable selection phase.}
          \label{fig:gamma_trend}
\end{figure}

Fit a regression model with a linear date term and a DOW effect to $\hat{\gamma}^*_t$, analogous to Equation \ref{eq:reg}. Variable selection is then performed. The fitted regression and predictions ($\hat{\gamma}^{\text{trend}}_t$) are shown in Figure \ref{fig:gamma_trend}.

\newpage
\subsubsection*{Step 6: Compute a Joint Probability Distribution over Tuning Parameters}
\begin{figure}[h!]
    \centering
	 \includegraphics[width=1\linewidth]{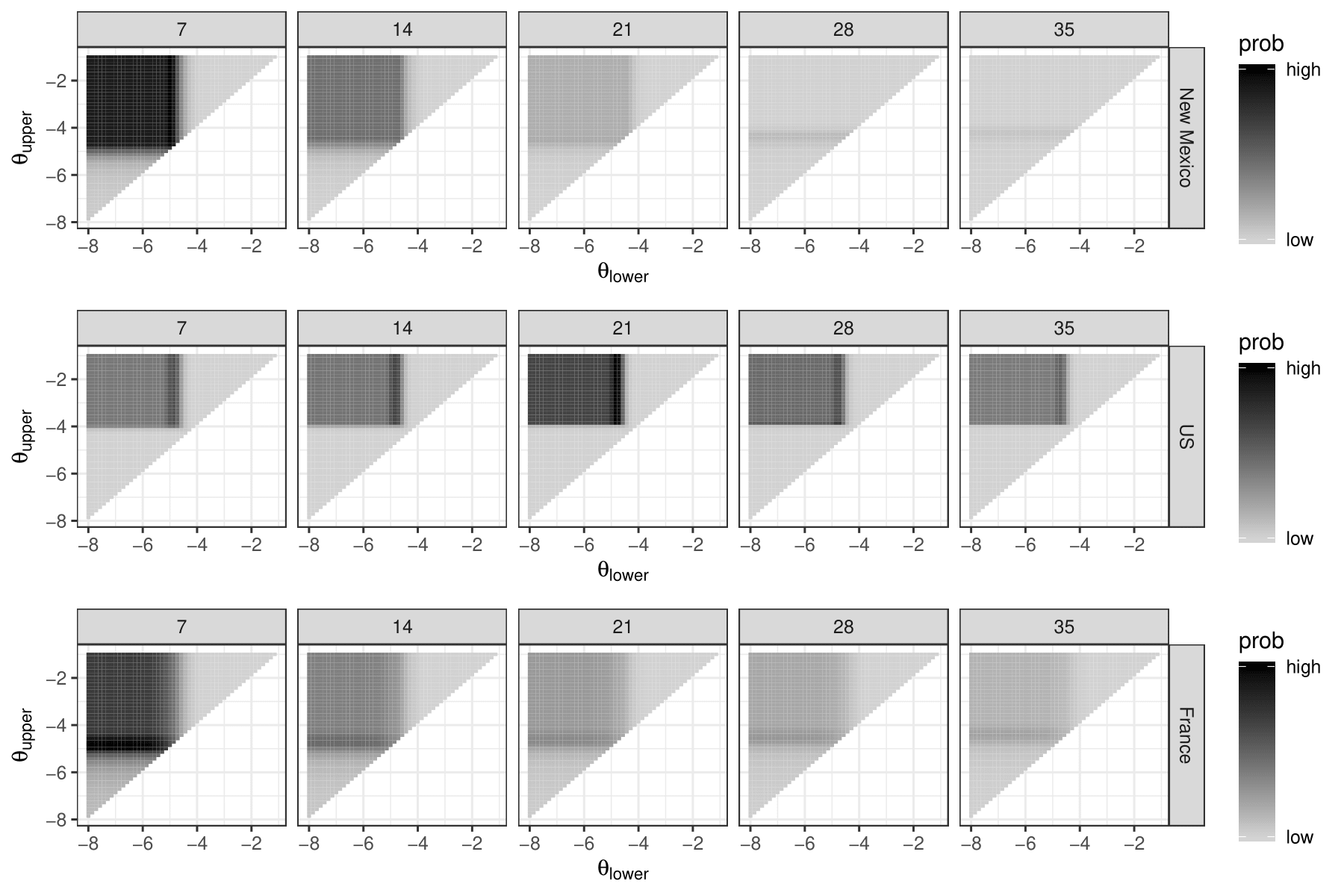}
          \caption{Joint probability distributions over tuning parameters $\nu$ (columns), $\theta_{\text{upper}}$, and $\theta_{\text{lower}}$ for New Mexico, the US, and France (rows). }
          \label{fig:deaths_joint_prob}
\end{figure}

The form of the forecasting model for $\hat{\gamma}_t^{\text{forecast}}$ is
\begin{align}
\label{eq:fcstform_gamma}
\hat{\gamma}_t^{\text{forecast}}(\nu,\theta_{\text{lower}},\theta_{\text{upper}}) &=
\begin{cases}
\theta_{\text{lower}} & \text{ if $\hat{\gamma}^{\text{trend}}_t < \theta_{\text{lower}}$}\\
\theta_{\text{upper}} & \text{ if $\hat{\gamma}^{\text{trend}}_t > \theta_{\text{upper}}$}\\
\hat{\gamma}^{\text{trend}}_t & \text{ otherwise}
\end{cases}
\end{align}
\noindent for $t = T^{\text{train}} + k$ and $k \in 1,2,\ldots, 14$. The parameters $\theta_{\text{lower}}$ and $\theta_{\text{upper}}$ act as a floor and a ceiling to $\hat{\gamma}^{\text{forecast}}_t$, keeping it from getting too large or too small. We evaluate $\hat{\gamma}^{\text{forecast}}_t$ on a grid over $\nu$, $\theta_{\text{lower}}$, $\theta_{\text{upper}}$ and compute the joint distribution as proportional to the inverse-distance between $\text{logit}^{-1}(\hat{\gamma}^{\text{forecast}}_t)$ and $\hat{\gamma}_t$, similar to Equation \ref{eq:invdist}. The estimated joint probability distribution over tuning parameters is shown in Figure \ref{fig:deaths_joint_prob}.

\subsubsection*{Step 7: Produce the Reported Deaths Forecast}
The final step is to simulate reported deaths. The purpose of the previous steps was to get a joint probability distribution over the tuning parameters that can be used to sample from. 

If more than 14 of the last 28 days had zero reported deaths, we take iid samples of future reported deaths from the empirical distribution of outlier adjusted reported deaths over the last 28 days. 
If no deaths were reported over the last 28 days, we sample future reported deaths as iid Bernoulli draws with success probability equal to 1/29. 
If 14 or more of the last 28 days observed at least 1 reported death, we simulate $\hat{\gamma}^{\text{forecast}}_{T+k}$ for $k \in {1,2,...,K}$ by doing the following:

\begin{enumerate}
\item Fit the regression outlined in Step 5 to days $T$ to $T - 27$. Use this to compute $\hat{\gamma}^{\text{trend}}_{T+k}$.
\item Draw a vector of $(\nu, \theta_{\text{lower}}, \theta_{\text{upper}})$ from the joint distribution computed in Step 6. 
\item Compute $\hat{\gamma}_{T+k}^{\text{forecast}}$ following Equation \ref{eq:fcstform_gamma}.
\item Compute logit$^{-1}(\hat{\gamma}_{T+k}^{\text{forecast}})$.
\item For $k=1,2,\ldots,K$, compute
\begin{enumerate}
\item $y^{\text{forecast}}_{d,T+k} = \text{logit}^{-1}(\hat{\gamma}^{\text{forecast}}_{T+k}) f(\boldsymbol{\delta}^{\text{forecast}}_{c,1:t},\nu)$, where $\delta^{\text{forecast}}_{c,t} = y_{c,t}$ if $k-\nu \leq 0$ and $\delta^{\text{forecast}}_{c,t} = y^{\text{forecast}}_{c,t}$ if $k-\nu > 0$.
\end{enumerate}
\end{enumerate}

\begin{figure}[h!]
    \centering
	 \includegraphics[width=1\linewidth]{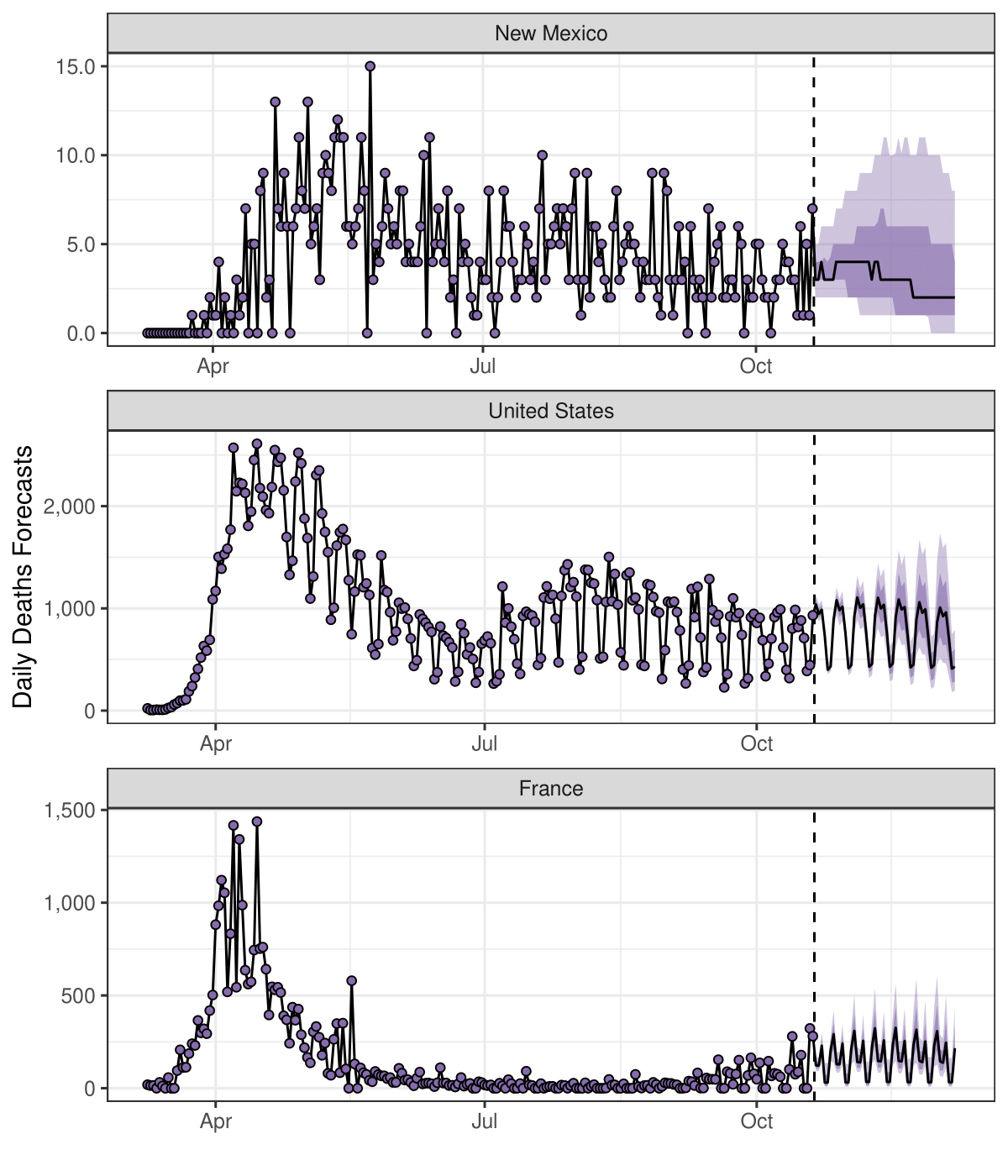}
          \caption{The median (black line) and 50\% and 80\% prediction intervals (ribbons) for New Mexico, the US, and France for daily reported deaths.}
          \label{fig:deaths_fcsts}
\end{figure}

Figure \ref{fig:deaths_fcsts} shows the daily deaths forecasts for New Mexico, the US, and France.

\newpage

\bibliographystyle{plain}
\bibliography{references}

\begin{thebibliography}{1}

\bibitem{jacquez1993stochastic}
{Jacquez, John A and Simon, Carl P}.
\newblock {The stochastic SI model with recruitment and deaths I. Comparison
  with the closed SIS model}.
\newblock {\em Mathematical biosciences}, 117(1-2):77--125, 1993.

\bibitem{tsoutliers}
{Javier López-de-Lacalle}.
\newblock {\em {tsoutliers: Detection of Outliers in Time Series}}, 2019.
\newblock R package version 0.6-8.

\bibitem{picard1990data}
{Picard, Richard R and Berk, Kenneth N}.
\newblock {Data splitting}.
\newblock {\em {The American Statistician}}, 44(2):140--147, 1990.

\end{thebibliography}

\end{document}